\documentclass[aps,prl,epsfig,showkeys,floats,twocolumn,amssymb,amsmath,floatfix]{revtex4-1}

\usepackage{graphicx}
\usepackage{xspace}
\usepackage{color}

\definecolor{insetbox}{rgb}{1,0.7,0}

\begin{document}

\title{A steering mechanism for phototaxis in {\em Chlamydomonas}}

\author{Rachel R. Bennett}
\author{Ramin Golestanian}
\email[]{ramin.golestanian@physics.ox.ac.uk}
\affiliation{Rudolf Peierls Centre for Theoretical Physics, University of Oxford, Oxford OX1 3NP, UK}

\date{December 11, 2014}

\begin{abstract}
{\it Chlamydomonas} shows both positive and negative phototaxis. It has a single eyespot near its equator and as the cell rotates during forward motion the light signal received by the eyespot varies. We use a simple mechanical model of {\it Chlamydomonas} that couples the flagellar beat pattern to the light intensity at the eyespot to demonstrate a mechanism for phototactic steering that is consistent with observations. The direction of phototaxis is controlled by a parameter in our model and the steering mechanism is robust to noise. Our model shows switching between directed phototaxis when the light is on and run-and-tumble behaviour in the dark.

\end{abstract}


\maketitle

Physical models are useful for understanding complex biological processes. For example, Berg and Purcell described how a microorganism could compare the concentration of chemoattractants over a time interval of a few seconds \cite{BergChemotaxis1}. Subsequent experiments demonstrated that wild-type {\it E. coli} cells perform such comparisons \cite{BergChemotaxis2,BergChemotaxis3}. {\it E. coli} swims with run-and-tumble motion and the cell uses the temporal concentration comparisons to modify the time between tumble events so that runs are longer when the cell swims up the concentration gradient. Friedrich and J\"{u}licher described how larger microorganisms that swim on noisy helical paths can navigate in a concentration gradient using a simple feedback mechanism \cite{FriedrichHelicalChemo}.


In phototaxis, organisms move towards or away from a light source. The flux of light is a vector quantity so the orientation of an organism determines the measured  intensity, whereas chemotaxing organisms measure a scalar quantity. Observations show that {\it Chlamydomonas} exhibits run-and-tumble in the dark \cite{PolinChlamyRT}, but we do not know of any observations of run-and-tumble motion during phototaxis. How can {\it Chlamydomonas} use run-and-tumble as a search strategy in darkness, and stop tumbles and steer directly towards or away from a light source? A model of {\it Chlamydomonas} also needs to swim at low Reynolds number, which is non-trivial \cite{PurcellLowRe}, and have synchronized flagella during runs \cite{PolinChlamyRT}. It would be remarkable to bundle all these complex engineering requirements into one simple model. We develop a simple model with coupling to the light intensity at the eyespot and intrinsic noise to investigate how {\it Chlamydomonas} could produce these behaviours.

\begin{figure}
   \centering
   \includegraphics[trim=15mm 17cm 107mm 17mm, clip=true, width=\columnwidth]{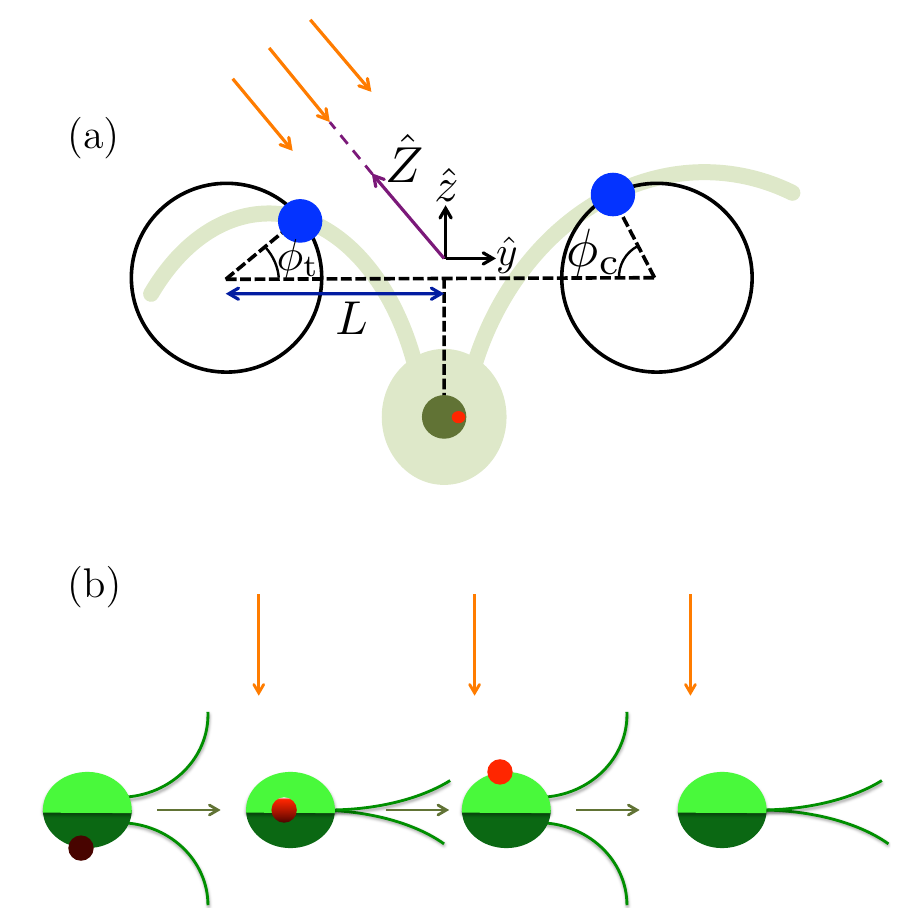}
   \caption{(Online version in colour.) (a) The three-sphere model: Each flagellum is represented by a sphere, shown in blue, moving around a circular trajectory and has phase $\phi_{\rm c,t}$. Each sphere is driven by a phase-dependent tangential force. A third sphere, shown in dark green, represents the cell body and the red spot indicates the position of the eyespot. The green underlay is a schematic of the real cell, and the orange arrows show incoming light. (b) The cell rotates about the $\hat{z}$-axis so when it swims perpendicular to the light direction (orange arrows), the light signal varies. For half the cycle the eyespot is in the dark, then when the eyespot crosses the horizon line the light signal varies as $\cos{\vartheta}$, where $\vartheta$ is the angle between the outward normal from the eyespot and the direction towards light.}
   \label{fig:3SM}
\end{figure}

{\it Chlamydomonas} is a photosynthetic unicellular alga that shows both positive and negative phototaxis. It detects light with an eyespot near the cell's equator, offset about $45^{\circ}$ ahead of the flagellar beat plane (the $y$-$z$ plane shown in figure \ref{fig:3SM}(a)). It is interesting that evolution has placed the eyespot on the side of the cell instead of the anterior or posterior. The equatorial position is advantageous because {\it Chlamydomonas} rotates about its body during forward swimming, allowing the eyespot to scan the incoming light from different directions \cite{FosterSmythAntennas,RufferNultschHighSpeed,HarzEyespotChange,SchallerKeepOrientation}. When the cell swims perpendicular to the incoming light, the signal varies sinusoidally; when the cell swims parallel to the light, the signal is constant.

Light excitation of the eyespot triggers photocurrents causing an influx of Ca$^{2+}$ ions to the flagella \cite{HarzHegemannPRC,WitmanRev,TammCa2+,HollandRhodopsin,HegemannReview}.
The response to the influx of Ca$^{2+}$ ions is different in the cis-flagellum (closest to the eyespot) and trans-flagellum (furthest from the eyespot). Kamiya and Witman studied axonemes in reactivated demembranated models and observed that increasing the Ca$^{2+}$ concentration decreases the beat amplitude of the cis-axoneme and decreasing the Ca$^{2+}$ concentration decreases the beat amplitude of the trans-flagellum \cite{KamiyaWitmanCaAxoneme}.

R\"{u}ffer and Nultsch performed high-speed cinematography photoresponse experiments on fixed cells to study the change in beat frequency and pattern in response to step-up and step-down white light stimuli \cite{RufferNultschPhoto}. They observed two step-up responses and two step-down responses; they suggest that the different responses to each type of stimulus correspond to positive and negative phototaxis. They observed that the cis and trans flagellum have the same frequency response, however, the beat amplitude of the cis and trans flagella changed in the opposite sense. For `type (+)' cells, on step-up stimuli the amplitude of the trans flagellum increases and the amplitude of the cis flagellum decreases; for `type (-)' cells, on step-up stimuli the amplitude of the trans flagellum decreases and the amplitude of the cis flagellum increases (vice versa for step-down stimuli). These results were confirmed by subsequent measurements with better sensitivity and allowed simultaneous measurements of photocurrents \cite{HollandOptoElec,JosefOptoElec}. A mutant strain {\it ptx}1 does not show phototaxis: the two flagella show the same amplitude response \cite{HorstWitmanptx1,RufferNultschptx1}.

The eyespot is located about $45^{\circ}$ ahead of the flagellar beat plane, so at typical 2 Hz rotation the eyespot passes some fixed point 60ms ahead of the beat plane.
The lag time between light stimuli and beat pattern changes are 30-40ms, suggesting that in free swimming cells, the beat pattern changes when the light direction is nearly in the beat plane \cite{RufferNultschHighSpeed,RufferNultschPhoto}. The sensitivity of the photoreceptor is optimised for this frequency of rotation \cite{RotationFreqYoshimura}. Schaller {\it et al} suggested that the $45^{\circ}$ offset of the eyespot is important for the cell keeping its correct orientation towards or away from the light \cite{SchallerKeepOrientation}. They showed that cells swim with helical motion so the eyespot either tilts towards the light or away from the light depending on whether the cis or trans flagellum has the larger beat amplitude and whether the cell is positively or negatively phototactic. They showed that single photons can cause directional changes and argued that in order to keep the correct orientation, the eyespot should be shaded by tilting away from the light.

The direction of phototaxis is affected by pre-irradiation, concentration of cations, the intensity of the directional light source and the wavelength of background monitoring light \cite{RufferNultschPhoto,CalciumSign,IntensitySign,PhotosynthesisModulation}. Ref. \cite{PhotosynthesisModulation} reports evidence that photosynthesis affects the sign of phototaxis. The mechanism that controls the sign of phototaxis is still an open question.

Drescher {\it et al.} developed a model to study phototaxis in {\it Volvox carteri}, a colonial alga with thousands of cells like {\it Chlamydomonas} on the surface of a spherical extracellular matrix \cite{DrescherVolvox}. Each individual cell responds to a step up in light by decreasing its flagellar activity, then after some time adapting to its baseline. There is no central nervous system, but the coordinated response gives phototactic steering. {\it Volvox} rotates about its swimming direction axis, and Drescher's model shows that this is essential for accurate phototaxis. Although {\it Chlamydomonas} and {\it Volvox} are closely related, their steering mechanisms for phototaxis are different: {\it Chlamydomonas} is a single cell that responds to a single stimulus, but {\it Volvox} has many individual cells each responding to a different stimulus (some cells are shaded and the intensity received by the illuminated cells depends on their angle to the light). The necessity of rotation is a common feature with our model, where the rotation of {\it Chlamydomonas} is necessary for its single eyespot to scan the light in different directions.

Here, we use a simple mechanical model to demonstrate how opposite changes in beat pattern can give phototactic steering. Previously we developed the model to study synchronization and emergent run-and-tumble behaviour \cite{BennettGolestanian}. Here we extend the model by coupling the light intensity at the eyespot to a parameter that describes the beat pattern.

\emph{The three-sphere model.}---
Simple actuated bead models have been used extensively to study swimming \cite{Najafi:2004,Avron:2005,DeSimone:2012,Leoni:2009,Tierno:2008a} and synchronization \cite{Vilfan:2006,Uchida:2010,Kotar:2010,Uchida:2011,Brumley:2012,Friedrich:2012} at low Reynolds number. Inspired by measurements of the flow field around {\it Chlamydomonas}, found to be well represented by three Stokeslets \cite{ChlamyFlowField,ChlamyFlowFieldT}, we represent the flagella as spheres moving around circular trajectories in opposite directions. A third sphere represents the cell body; the model is shown in figure \ref{fig:3SM}.
Each flagellar sphere is driven by a tangential force that is phase dependent, representing the beat pattern. We consider driving force profiles of the form $F(\phi_i)=F_0(1+a_i\sin{(n\phi_i+\Delta)})$,
where $\phi_i$ is the phase of flagellum $i={\rm c, t}$ and $\Delta$ is a constant phase.
Evidence for modelling the beat pattern using phase dependent forcing is shown by Geyer {\it et al.}, who develop a similar model and compare with detailed observations of the flagellar beat pattern of {\it Chlamydomonas} \cite{Geyer:2013}. 
The beat pattern is an essential feature in this model for producing the behaviour observed in {\it Chlamydomonas}, for example, synchronized beating \cite{BennettGolestanian}.

{\it Chlamydomonas} has a diameter of $\sim 10 \mu$m and swims in water with velocity $\sim 100 \mu$m${\rm s}^{-1}$, so the Reynolds number is $\sim 10^{-3}$. The Reynolds number is small so velocities are linearly related to forces and Stokes friction acts on each sphere. Hydrodynamic interactions between the spheres break the time reversal symmetry, which is a requirement for net propulsion at low Reynolds number. Further details of the model are presented in Ref. \cite{BennettGolestanian}, where we found emergent run-and-tumble behaviour when we added noise to the parameter $a_i$.

Here, we choose $F(\phi_i)=F_0(1+a_i\sin{(\phi_i+\Delta)})$ and couple the coefficient $a_i$ to the light intensity at the eyespot
\begin{equation}
   \begin{array}{r}
   a_{\rm c}=a_0+p\log{(1+I(t))}+\zeta_{\rm c}(t), \\
   a_{\rm t}=a_0-p\log{(1+I(t))}+\zeta_{\rm t}(t),
   \end{array}
   \label{eq:coupling}
\end{equation}
where $p$ controls the strength of coupling and sign($p$) determines the direction of phototaxis, and $I(t)$ is the intensity of light at the eyespot which varies as the cell rotates. The noise, $\zeta_i(t)$, has Gaussian probability distribution with zero mean and correlation function $\langle \zeta_i(t)\zeta_j(t')=\sigma^2\delta_{ij}\delta(t-t')\rangle$. The noise is not part of the steering mechanism in the presence of light, but we include it to show that steering is robust in the presence of noise. In the dark, the noise allows the cell to show run-and-tumble motion.
We also include a phase shift when there is no light:
$\Delta=0$ if $I>0$ and $\Delta=\pi/2$ if $I=0$.
We choose $a_0=0.6$ so that we have stable synchronized beating in the absence of the coupling term. The cis and trans coupling terms in equation (\ref{eq:coupling}) have opposite signs, motivated by observations that the beat amplitudes of the flagella respond in opposite ways to light stimuli \cite{RufferNultschPhoto}. We couple the light logarithmically to allow responses over a larger range of light intensity.

We denote the axes in the lab reference frame $(\hat{X},\hat{Y},\hat{Z})$ and choose the directional light to be in the $-\hat{Z}$-direction with intensity $I_0$. We denote the axes in the rotating cell frame $(\hat{x},\hat{y},\hat{z})$ where $\hat{z}$ points in the swimming direction and ${\hat{y}}$ points towards the cis flagellum. The eyespot points in the direction $(\hat{x}+\hat{y})/\sqrt{2}$, so approximating the cell body as opaque, the light intensity at the eyespot is
$I(t)=I_b+I_0 \hat{Z}\cdot(\hat{x}+\hat{y}) H\big(\hat{Z}\cdot(\hat{x}+\hat{y})\big)/\sqrt{2},$
where $I_b$ is isotropic background light and $H$ is the Heaviside step function. The intensities of the background light $I_b$ and the directional light $I_0$ are chosen such that $|a_{\rm c, t}|<1$.

The angular velocity about the $x$-axis that this model produces can be approximated as
\begin{equation}
   \omega_x=\omega_0(a_{\rm t}-a_{\rm c})+\omega_1 \tanh{(K(a_{\rm t}-a_{\rm c}))}\cos{(\Omega_b t)}
   \label{eq:omx}
\end{equation}
where $K\approx 100$ and $\Omega_b$ is the flagellar beat frequency. We use the full model and not approximation (\ref{eq:omx}) in the results presented here. The observed rotation about the $\hat{z}$-axis, $\Omega_z\approx \Omega_b/20$, can be produced by the model when we tilt the circular trajectories of the flagellar spheres out of the $y$-$z$ plane in opposite directions. The forward motion produced by the model is oscillatory and can be approximated by, $u = u_0+u_1 \cos(\Omega_b t)$, in agreement with experiments \cite{RaceyChlamySpeed}. The initial orientation angle $\theta_0$ is the angle between the cell's orientation $\hat{z}$ and the direction towards the light $\hat{Z}$. We choose an initial condition such that the $y$-$z$ plane lies in the $Y$-$Z$ plane.

\begin{figure*}
   \centering
   \includegraphics[trim=2cm 40mm 22mm 17mm, clip=true, width=2\columnwidth]{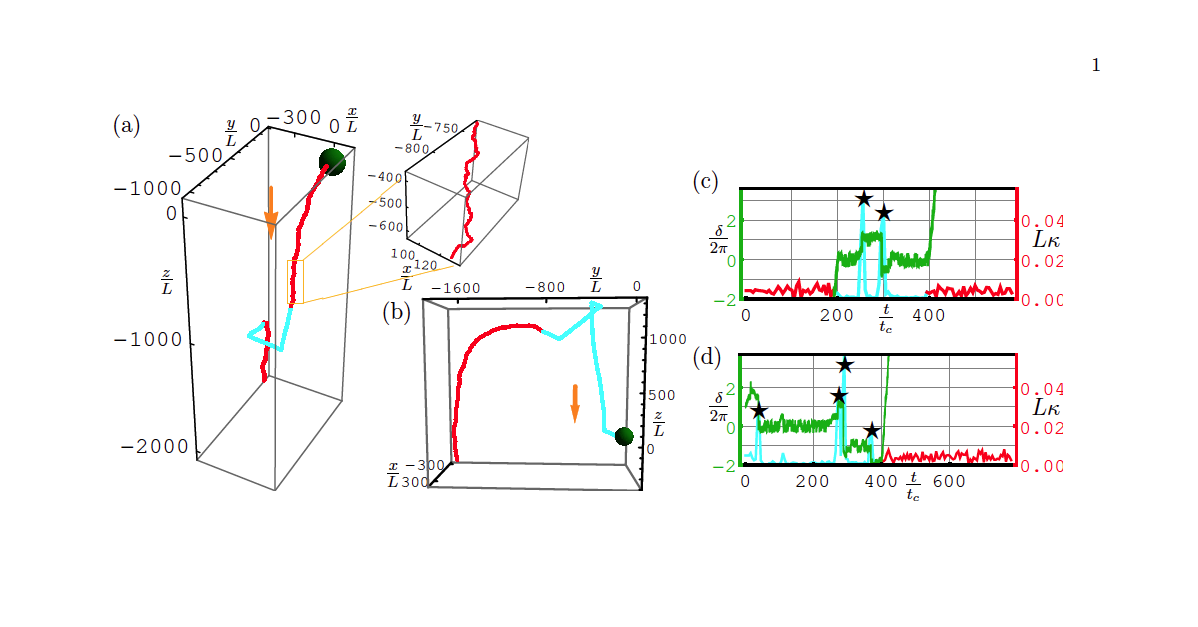} 
   \caption{(Online version in colour.) Example trajectories of the cell with alternating between light and dark shown in terms of length scale $L$, where $2L$ is the average distance between flagella beads, and time scale $t_c=6\pi\eta a L/F_0$, where $\eta$ is viscosity of water and $a$ is radius of the beads. The parameters used are $p=-1$, $I_b=0.1$, $I_0=0.1$, $\sigma=4.7 {\rm x} 10^{-3}$ and the initial condition is $\theta_0=\pi/2$. (a) For $0<t<T/3$ and $2/3 T < t < T$ the light is on (red), where $T$ is the total time of trajectory. For $T/3<t<2/3 T$ the cell is in the dark (blue) and shows run-and-tumble behaviour. The orange arrow shows the direction of the incoming light and the green spot (not to scale) shows the initial position of the cell. The inset shows the helical shape of the trajectory. (b) An example trajectory where the light is off for $t<T/2$ and the light is on for $t>T/2$. The cell is in the dark for twice as long as in (a) and we see twice as many tumbles. (c, d) (Red/blue) Curvature, $\kappa$, of trajectory shown in (a) and (b) respectively, when we average over the helical motion. The large peaks in the curvature during darkness correspond to tumbles. (Green) Phase difference, $\delta$, of the beating flagella during darkness. The black stars highlight the peaks in curvature.}
   \label{fig:3SMtraj}
\end{figure*}

\emph{Results.}---
The three-sphere model shows positive phototaxis when $p>0$ and negative phototaxis when $p<0$ for all initial orientations, $\theta_0$, and choices of $I_b$, $I_0$ that satisfy $|a_{\rm c, t}|<1$. Even if we start the cell on a trajectory directly away from the light, if $p=1$ then the negative phototaxis direction is unstable and the steering mechanism turns the cell towards the light. Similarly, when $p=-1$, if we start the cell on a trajectory towards the light then the steering mechanism turns the cell away from the light.
On average the cell travels on a helical path as shown in the inset of figure \ref{fig:3SMtraj}(a).

The steering mechanism is robust to noise; when the light is on the cell moves towards or away from the light, and if a tumble occurs then the cell quickly reorients itself towards or away from the light again. We have tested noise strengths $\sigma \in [10^{-4}, 5 {\rm x} 10^{-3}]$ and we find that the cell steers directly towards or away from the light with only an occasional tumble that is quickly corrected for the larger noise strengths. Figure \ref{fig:3SMtraj} shows trajectories with $p=-1$ where the cell alternates between light and dark phases. The cell steers away from the light, then when the light is switched off the cell shows run-and-tumble behaviour. When the light is turned on again the cell immediately steers away from the light again. If $\sigma < 10^{-4}$ then we see the same dynamics in the light, but in the dark the run duration becomes longer and we would need longer simulation times to see tumble events. In Ref. \cite{BennettGolestanian} we show that the characteristic run duration varies as $1/\sigma$.

If the cell swims in periodic light with a period that is shorter than the characteristic run duration and longer than rotation period, then the probability of a tumble event occurring during the dark phase is small and the cell can follow its path towards the light without interruption. The initial orienting of the cell towards or away from the light can be slowed by dark phases, but once the cell has reached the correct orientation, short periodic dark phases do not disturb the motion towards or away from the light.

Figures \ref{fig:3SMtraj}(c,d) show the curvature of the trajectories in figures \ref{fig:3SMtraj}(a,b) when we average over the helical motion and small fluctuations. We see that in the red sections when the light is on the curvature is small. During darkness we see peaks in the curvature that correspond to the tumbles and sections of very small curvature that correspond to straight runs. In Ref. \cite{BennettGolestanian} we found that when tumbles occur we also see jumps in the phase difference between the two flagella. The green line in figures \ref{fig:3SMtraj}(c,d) shows the phase difference and we see that the peaks in curvature correspond to jumps in phase difference, confirming that the peaks in curvature are tumbles and that we have run-and-tumble motion in the dark.

\begin{figure}
   \includegraphics[trim=20mm 203mm 110mm 17mm, clip=true, width=\columnwidth]{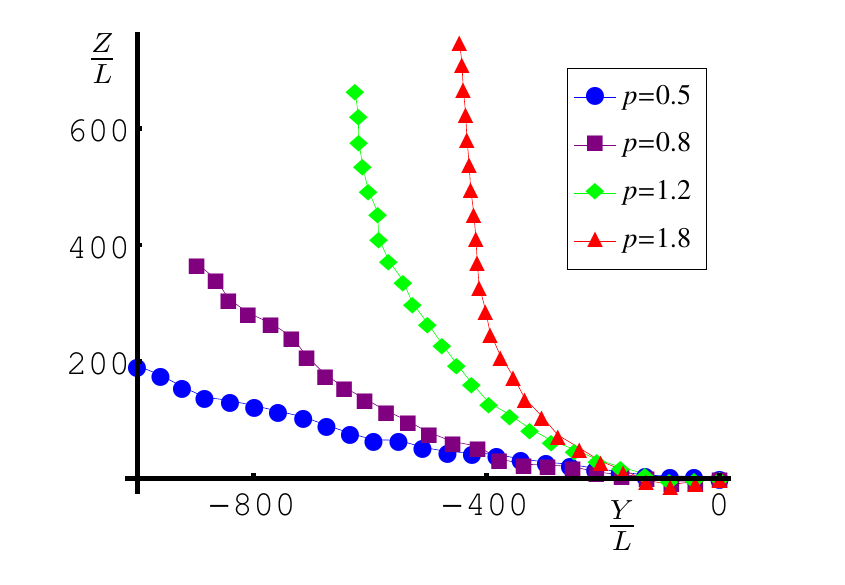}
   \caption{Trajectories of the cell with different values of coupling strength $p$ projected into the $Y$-$Z$ plane, where $t/t_c \in [0,180]$. Blue circles: $p=0.5$; purple squares: $p=0.8$; green diamonds: $p=1.2$; red triangle: $p=1.8$. The cell is initially oriented in the $-\hat{y}$ direction and steers towards the light coming in from the $\hat{Z}$ direction. At $t/t_c=180$, the cells with $p=1.2$ and $p=1.8$ have oriented towards the light, but the cells with $p=0.5$ and $p=0.8$ are still slowly bending towards the light.}
   \label{fig:steerP}
\end{figure}

Stronger coupling gives faster steering. This is shown in figure \ref{fig:steerP}, where the initial curvature towards the light is greater for greater values of $p$. Larger values of $p$ increase the mismatch between the cis and trans beat amplitudes which allows faster rotations in the beat plane.

\emph{Understanding the steering mechanism.}---
When we average over the fast oscillations within a flagellar beat cycle, the cell swims along a helical path and we consider how the helix bends towards or away from the light. The cell swims with linear velocity $u$ in the $\hat{z}$-direction and angular velocities $-\Omega_z$ about the $\hat{z}$-axis, $\Omega_z>0$, and $\omega_x$ about the $\hat{x}$-axis. This combination causes the cell to swim along a left-handed helix with radius $r=u|\omega_x|/\Omega_z^2$ and pitch length $\lambda=2\pi u/\Omega_z$. The angle between the cell trajectory and the helix axis is
\begin{equation}
   \gamma = \tan^{-1}{(2\pi r/\lambda)}=\tan^{-1}{(|\omega_x|/\Omega_z)}.
   \label{eq:gamma}
\end{equation}
The $\hat{y}$-axis points outwards in the radial direction when $\omega_x>0$ and points inwards in the radial direction when $\omega_x<0$. We can approximate the $x$-component of angular velocity using equation (\ref{eq:omx}) and average over the fast oscillations:
\begin{equation}
   \omega_x \approx -2p \omega_0 \log{(1+I(t))}.
   \label{eq:omxapprox}
\end{equation}

\begin{figure}
   \centering
   \includegraphics[width=\columnwidth]{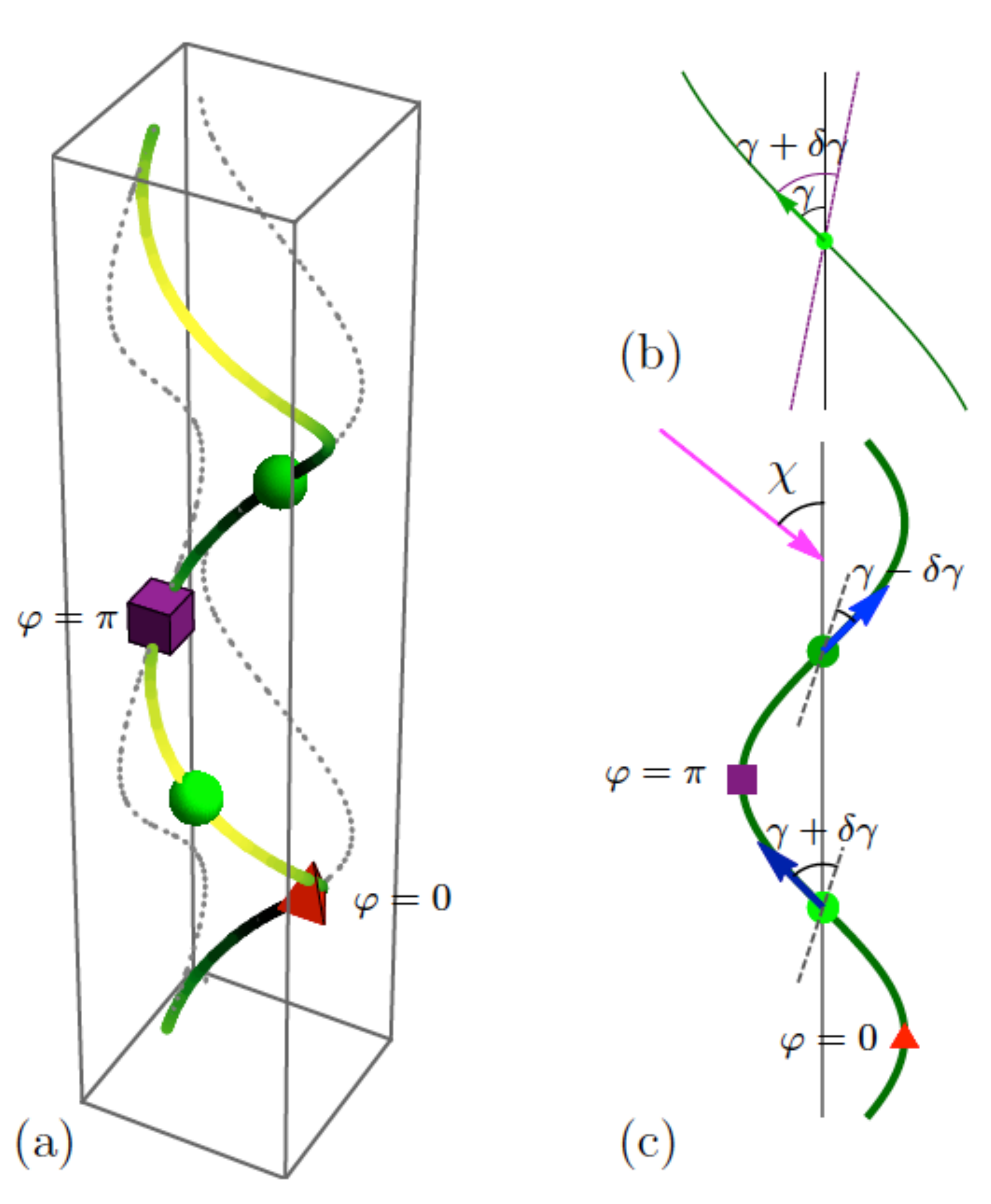}
   \caption{(Online version in colour.) (a) Diagram of helix showing phases corresponding to those marked in (c). (b) A change in $\gamma$ is facilitated by the helix axis changing direction by $\delta \gamma$. The purple dashed line shows the new cell axis after $\gamma \to \gamma+\delta\gamma$. (c) The effect of changes in $\gamma$ on the angle $\chi$ depend on the phase of the cell: $\delta \chi = \delta\gamma \sin{\varphi}$. The red triangle and purple square mark the positions $\varphi=0$ and $\varphi=\pi$, respectively. When the cell has position $\varphi \in (0,\pi)$, an increase (decrease) in $\gamma$ causes the helix axis to bend away from (towards) the light and a corresponding decrease (increase) in $\gamma$ when $\varphi \in (\pi,2\pi)$ also causes the helix axis to bend away from (towards) the light. The incoming light is marked by the pink arrow and the blue arrows show the direction that the cell travels along the helix. The dashed lines at $\varphi=\pi/2$ and $\varphi=3\pi/2$ show the new direction of the helix axis if $\gamma$ increases and decreases, respectively, by $\delta \gamma$ at these positions.}
   \label{fig:helixsteer}
\end{figure}

For either choice of $p$, from equations (\ref{eq:gamma}) and (\ref{eq:omxapprox}) we see that when $I(t)$ decreases (increases), then $|\omega_x|$, and hence $\gamma$, decreases (increases). To see how changes in the pitch angle $\gamma$ affects steering, we consider the angle between the helix axis and the $\hat{Z}$-axis (light direction) which we denote $\chi$. If the helix angle $\gamma$ changes as a result of the changing light intensity, there are two possible ways that the correct angle can be made: either the cell changes its trajectory by rotating about its $\hat{y}$-axis or the helix axis changes direction. But since the flagella beat in the $y-z$ plane, the cell does not generate the required rotation about the $\hat{y}$-axis. Therefore, the change in $\gamma$ is produced by a change in direction of the helix axis, as shown in figure \ref{fig:helixsteer}(a). This change in direction changes the angle $\chi$ between the light direction and the helix axis.

Increasing and decreasing light produce opposite changes in $\gamma$, but since these changes happen when the cell is at different positions along the helix, the opposite changes in $\gamma$ can work together to produce the same change in $\chi$.
Let $\varphi \in [0, 2\pi)$ be the phase of the cell on the helix where $\varphi=0$ is the position furthest from the incoming light, marked with a red triangle in figure \ref{fig:helixsteer}(c). When $\varphi \in (0,\pi)$, an increase (decrease) in $\gamma$ causes the helix axis to bend away from (towards) the light and a decrease (increase) in $\gamma$ when $\varphi \in (\pi,2\pi)$ also causes the helix axis to bend away from (towards) the light.

We consider what happens for each choice of sign($p$). First we consider $p>0$. In this case, $\omega_x<0$ so the eyespot moves around the inside of the helix. As the cell moves from $\varphi=0$ to $\varphi=\pi$, on average the cell sees a decrease in the light intensity as the cell body moves around to shield the eyespot, so $\gamma$ decreases, causing the helix axis to bend towards the light. Similarly, as the cell moves from $\varphi=\pi$ to $\varphi=2\pi$, on average the cell sees an increase in the light intensity so $\gamma$ increases, also resulting in the helix axis bending towards the light. Therefore we have positive phototaxis for $p>0$.

If $p<0$, then $\omega_x>0$ and the eyespot moves around the outside of the helix. When the cell moves from $\varphi=0$ to $\varphi=\pi$ the light intensity at the eyespot increases, $\gamma$ increases, and the helix axis bends away from the light. When the cells moves from $\varphi=\pi$ to $\varphi=2\pi$, the light intensity decreases, $\gamma$ decreases, and the helix axis bends away from the light. Therefore we have negative phototaxis for $p<0$.

The direction of phototaxis is controlled by sign($p$). If we use linear coupling instead of the logarithmic coupling used in equation (\ref{eq:coupling})---or indeed any other monotonically increasing functional form---then we find the same phototaxis directions for each choice of sign($p$) and the mechanism is the same as that described above, provided the coupling strength does not exceed the permissible range for the $a_i$ coefficients.

The rate that the cell turns towards or away from the light depends on $|p|$. The rate of bending towards or away from the light is $\dot{\chi} = \dot{\gamma}\sin{\varphi}$, where
\begin{equation}
   \dot{\gamma} \approx \frac{2|p|\omega_0\dot{I}}{\Omega_z\big(1+I(t)\big)}.
   \label{eq:gamdot}
\end{equation}
$\dot{I}$ changes sign at approximately the same time as $\sin{\varphi}$, so increasing $|p|$ increases the rate at which $\chi$ decreases or increases towards 0 or $\pi$, respectively.

\emph{Discussion.}---
We have shown that in the three-sphere model of {\it Chlamydomonas} our simple coupling of the amplitude in the flagellar driving force to the intensity at the eyespot leads to phototactic steering. The direction of phototaxis and steering rate is controlled by the parameter $p$. As the cell rotates about its body axis, the light intensity viewed by the eyespot varies and this produces variation in the angular velocity. The cell swims along a helical path and variation in the angular velocity causes the helix axis to bend. The position of the cell along the helix affects whether the helix axis bends towards or away from the light so opposite changes in light intensity that occur on opposite sides of the helix can work together to bend the cell towards or away from the light.

We have included noise in our model and shown that the steering mechanism is robust with respect to the noise. When the light is on, the cell steers towards or away from the light and travels along a helical path.
In darkness, the cell shows run-and-tumble behaviour and the phase dependence in our model allows us to control the synchronization properties of the flagella, as described in Ref. \cite{BennettGolestanian}. The model includes a phase shift in the beat pattern between light and darkness and this gear change enables the cell to run-and-tumble effectively. Hydrodynamic interactions allow the model to break time reversal symmetry; a requirement for swimming at low Reynolds number. Therefore, the simple model successfully meets our complex engineering needs of low Reynolds swimming, synchronized beating, run-and-tumble in darkness and phototaxis.

In periodic light with a period that is shorter than the characteristic run duration, there are very few tumbles and our model maintains its path towards or away from the light in the same way as it does with a constant light source, although the initial orienting time can be longer. It would be interesting to see if real {\it Chlamydomonas} cells phototax as effectively in periodic light as they do in constant light when the period is shorter than the typical run duration and longer than the rotation period about the body axis.

We do not vary the parameter $a_0$ because the stability of the synchronized state is sensitive to this parameter as discussed in Ref. \cite{BennettGolestanian}. We only get the desired behaviour of phototactic steering in the light and run-and-tumble in the dark for a narrow range of $a_0$. The phase dependence of the driving force is essential for the model to produce the different types of observed behaviour.

The direction of phototaxis is controlled by $p$ in our model. {\it Chlamydomonas} usually shows positive phototaxis in low to moderate light intensities and negative phototaxis in high light intensity, so $p=2H(I_b+I_0-I_c(\lambda))-1$ is a reasonable choice for a basic model, where the critical light intensity $I_c(\lambda)$ depends on the wavelength of both background and directional light.

The magnitude of the coupling parameter $|p|$ increases the rate of steering towards or away from the light. A quick glance at equation (\ref{eq:gamdot}) suggests that increasing (decreasing) the spinning frequency, $\Omega_z$, decreases (increases) the rate of steering, $\dot{\chi}$. However, increasing (decreasing) $\Omega_z$ also increases (decreases) the rate of change of viewed light intensity, $\dot{I}$, which increases (decreases) the steering rate. Our numerical results do not show significant variation of the steering rate with $\Omega_z$, when the spinning frequency is at least an order of magnitude slower than the beat frequency. If the spinning and beat frequencies become comparable then we get nonlinear terms in equation (\ref{eq:gamdot}).


\emph{Acknowledgements.}---
We would like to thank the EPSRC for financial support.

\end{document}